\documentclass[aps,reprint,showpacs,superscriptaddress]{revtex4-1} 

\usepackage{hyperref} 
\hypersetup
{
    colorlinks=true,       
    linkcolor=blue,        
    citecolor=blue,        
    urlcolor=blue           
}

\usepackage{amsmath}
\usepackage{amssymb}
\usepackage{bm}
\usepackage{graphicx} 
\usepackage{subfigure} 
\usepackage{dcolumn} 
\usepackage{color}
\usepackage{ulem}

\let\oldhat\hat
\renewcommand{\hat}[1]{\oldhat{\mathbf{#1}}} 
\renewcommand{\vec}[1]{\mathbf{#1}}

\let\revappendix\appendix

\begin{document}
\title{Nonreciprocal spin waves in a chiral antiferromagnet without the Dzyaloshinskii-Moriya interaction}
\author{Suik Cheon}
\author{Hyun-Woo Lee}
\email{hwl@postech.ac.kr}
\affiliation{Department of Physics, Pohang University of Science and Technology, Pohang 37673, Korea}
\author{Sang-Wook Cheong}
\affiliation{Rutgers Center for Emergent Materials and Department of Physics and Astronomy, Rutgers University, Piscataway, New Jersey 08854, USA} 

\date{\today}

\begin{abstract}
Non-reciprocal spin wave can facilitate the realization of spin wave logic devices.
It has been demonstrated that the non-reciprocity can emerge when an external magnetic field is applied to chiral magnets whose spin structures depend crucially on an asymmetric exchange interaction, that is, the Dzyaloshinskii-Moriya interaction (DMI).
Here we demonstrate that the non-reciprocity can arise even without the DMI.
We demonstrate this idea for the chiral antiferromagnet Ba$_2$NbFe$_3$Si$_2$O$_{14}$ whose DMI is very small and chiral spin structure arises mainly from the competition between symmetric exchange interactions.
We show that when an external magnetic field is applied, asymmetric energy gap shift occurs and the spin wave becomes non-reciprocal from the competition between symmetric exchange interactions and the external magnetic field.
\end{abstract}


\maketitle


\section{Introduction}
 
\begin{figure}[t!]
     \subfigure{\label{fig:helix}}
     \subfigure{\label{fig:cycloid}}
  \includegraphics[width=8.5cm]{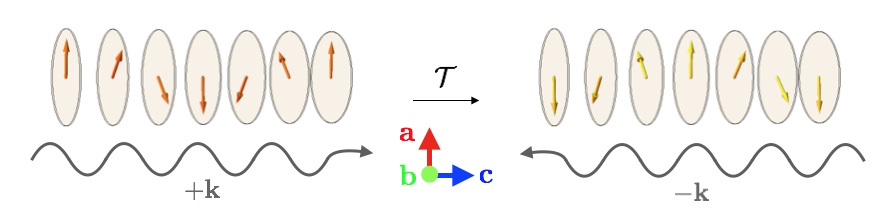}
\caption
{
Schematic illustration of helical chiral magnet.
The spin wave propagates along the $c$ axis (gray arrows), and orange arrows are time-reversal counterparts of yellow ones.
}
\label{fig:chiral}
\end{figure}

In many physical systems, waves propagating in opposite directions share the same characteristics. 
In certain special systems, on the other other hand, waves propagating in opposite directions may exhibit different characteristics. 
For instance, waves with wave vector $\pm {\bf k}$ may have different frequencies.
Such non-reciprocity may endow functionalities which are difficult to realize in reciprocal systems. 
In particular, it was suggested~\cite{Chumak:2015fa,Cheong:2018kg} that spin wave non-reciprocity can facilitate the realization of spin wave logic devices, such as a spin current diode.
For the spin wave non-reciprocity to emerge in chiral magnets, certain symmetries should be broken.
In case of the chiral magnets depicted in Fig.~\ref{fig:chiral}, their spin Hamiltonians may be invariant under the time-reversal operation $\mathcal{T}$~\cite{Kataoka:JPSJ1987}, which enforces the spin wave dispersion relation to be reciprocal, $E({\bf k})=E(-{\bf k})$. 
Thus the time-reversal symmetry should be broken by some means to induce the non-reciprocity.

Recently, the spin wave non-reciprocity in chiral magnets has been studied.
The experimental results in chiral ferromagnets Cu$_2$OSeO$_3$~\cite{PhysRevB.93.235131}, MnSi~\cite{PhysRevB.94.144420}, FeGe and Co-Zn-Mn alloys~\cite{PhysRevB.95.220406} indicate that the non-reciprocity arises in these noncentrosymmetric systems when an external magnetic field is applied.
When the field direction is reversed, the sign of the non-reciprocity is also reversed.
In Cu$_2$OSeO$_3$, it was demonstrated~\cite{PhysRevB.93.235131} that the sign of non-reciprocity depends not only on the field direction but also on the sign of crystal chirality. 
Non-reciprocal spin wave dispersion relation has been reported for chiral antiferromagnet $\alpha$-Cu$_2$V$_2$O$_7$~\cite{PhysRevLett.119.047201} as well, for which, similar to chiral ferromagnets, the breakings of the time reversal and the spatial inversion symmetries are important.
We remark that in these examples~\cite{PhysRevB.93.235131,PhysRevB.94.144420,PhysRevB.95.220406,PhysRevLett.119.047201}, the very existence of the chiral magnetism relies crucially on the Dzyaloshinskii-Moriya interaction (DMI)~\cite{PhysRev.120.91}.
Considering that the DMI itself requires some symmetries to be broken, it is natural in some sense to expect the spin wave dispersions to be non-reciprocal in these systems.

%
%
%
%
%

In this paper, we examine theoretically the spin wave dispersion in a chiral antiferromagnet Ba$_2$NbFe$_3$Si$_2$O$_{14}$ (BNFS) whose spin configuration forms the triangular-helical chiral magnetic order [Fig.~\ref{fig:spinconf}]. 
This system differs from the aforementioned chiral (anti)ferromagnets in that its chiral magnetic structure arises from the competition of symmetric exchange interactions~\cite{Kataoka:JPSJ1987} instead of the DMI.
We demonstrate that even without the DMI, the spin wave dispersion along the $c$-axis of BNFS becomes non-reciprocal and exhibits asymmetric energy shift when an external magnetic field is applied along the $c$-axis (parallel to $\vec{k}$). 
In contrast, in the chiral antiferromagnet $\alpha$-Cu$_2$V$_2$O$_7$~\cite{PhysRevLett.119.047201} whose spin configuration relies crucially on the DMI, the asymmetric energy shift appears when an external magnetic field is applied and is perpendicular to ${\bf k}$. 
We remark that in view of Ref.~\cite{Cheong:2018kg}, which examines possible non-reciprocity based on symmetry considerations (or symmetry-operational equivalence), the non-reciprocal spin waves in BNFS (our work) and $\alpha$-Cu$_2$V$_2$O$_7$~\cite{PhysRevLett.119.047201} correspond to two distinct cases (Figs. 3(c) and 3(d) of Ref.~\cite{Cheong:2018kg}, respectively), where the non-reciprocity is allowed by symmetries.
%
%
In addition, we mention that there is a distinct difference between non-reciprocal spin wave propagation and non-reciprocal light propagation~\cite{0034-4885-67-5-R03,PhysRevB.87.014421}; Light wave propagates with polarization, but spin wave has no polarization. 
Due to this difference, the physics of the spin wave non-reciprocity differs from the physics of the light non-reciprocity. 

The paper is organized as follows.
In Section~\ref{sec:model}, we introduce the spin Hamiltonian for BNFS with the external magnetic field, and obtain low energy spin wave excitations by the Holstein-Primakoff transformation.
In Section~\ref{sec:result}, we provide our numerical calculation results, and discussions.
Finally, the paper is summarized in Section~\ref{sec:summary}.
The detailed form of Hamiltonian is given in Appendix.

\section{\label{sec:model}Model Hamiltonian}

\begin{figure}[t!]
     \subfigure{\label{fig:crystal}}
     \subfigure{\label{fig:spinconf}}
  \includegraphics[width=8.5cm]{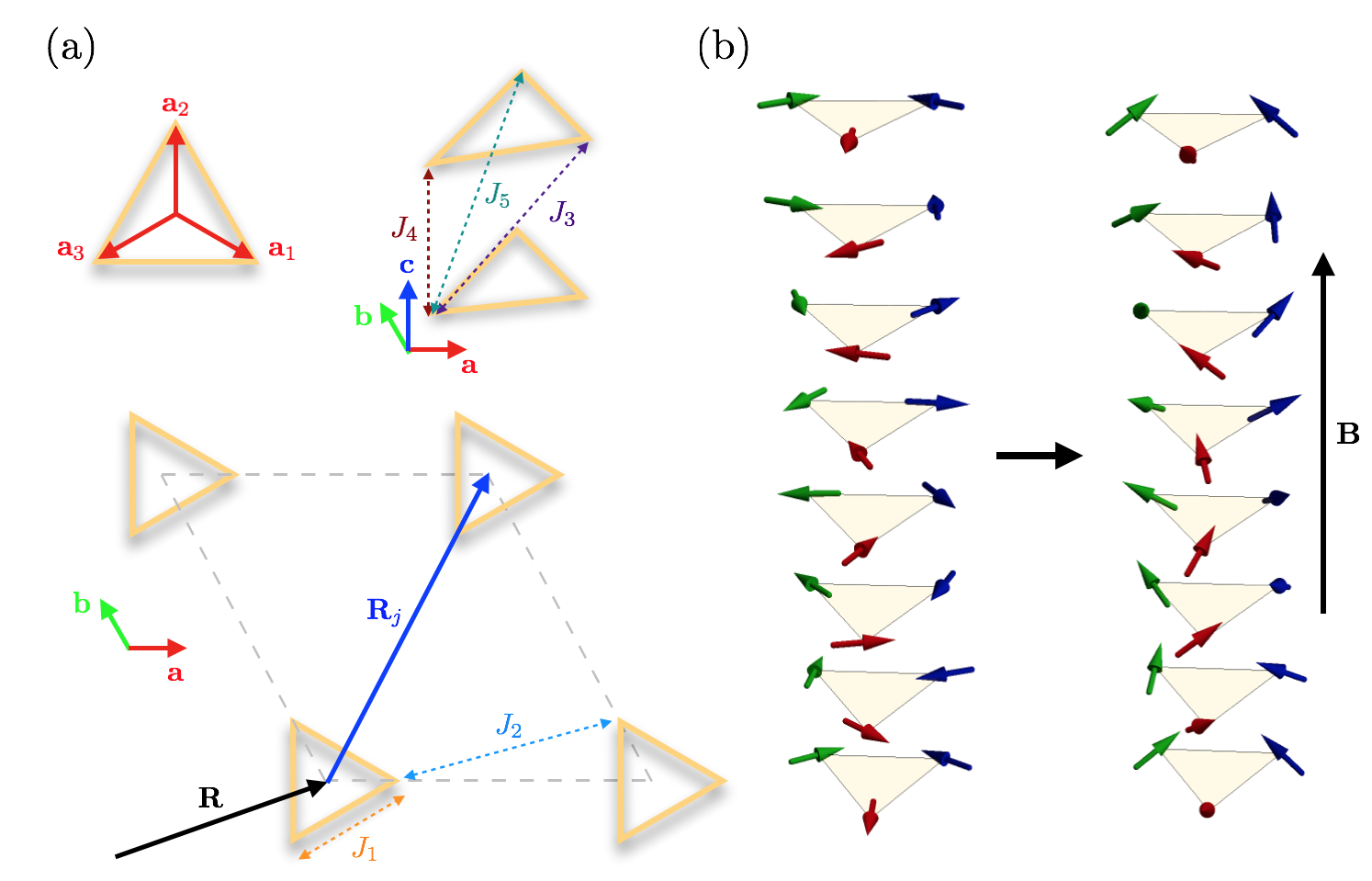}
\caption
{
(a) Magnetic exchange paths (dashed arrows) in Ba$_2$NbFe$_3$Si$_2$O$_{14}$ along the $c$ axis (top), and in the $ab$ plane (bottom). We only depict Fe$^{3+}$ triangles. The position vectors ${\bf a}_j$ and ${\bf R}_j$ are satisfied with $\sum_{j=1}^{3} {\bf a}_j = 0$ and $\sum_{j=1}^{3} {\bf R}_j = 0$, respectively. ${\bf R}$ indicates a reference point. (b) (Left) Equilibrium spin configuration in BNFS with ${\bf B}=0$ is helical spin arrangement. (Right) An external magnetic field ${\bf B}$ is applied along the $c$ axis. The ground state spin configuration in the presence of ${\bf B}$ is conical spin arrangement. 
}
\end{figure}

In order to obtain spin wave dispersion, we start with spin Hamiltonian for BNFS \cite{PhysRevLett.101.247201,PhysRevB.82.132408,PhysRevB.83.104426,PhysRevLett.106.207201,Simonet2012}.
This material crystalizes in noncentrosymmetric trigonal space group $P321$.
Figure~\ref{fig:crystal} shows that magnetic Fe$^{3+}$ ions in BNFS with $S = 5/2$ form a triangle lattice on $ab$ plane, and the exchange paths $J_1 \sim J_5$ are presented.
Below $T_N = 27$ K, the magnetic spin order occurs as shown in Fig.~\ref{fig:spinconf} \cite{PhysRevLett.101.247201}.
The magnetic spin order within each triangle follows $120^\circ$ arrangement, and spin arrangement is identical in all triangles within the same plane.
But along the $c$-axis, spin arrangement gets progressively tilted and forms a spin helix whose period is about $7$ layers.
BNFS has two kinds of chiralities. 
One is the helical chirality $(\epsilon_H=\pm 1)$, which represents the helical winding direction of spin as one moves along the $c$ axis, and the other is the triangular chirality $(\epsilon_T=\pm 1)$, which represents the winding direction of spin within each triangle.
The neutron scattering study~\cite{PhysRevLett.106.207201} on BNFS reports $\epsilon_H=1$ and $\epsilon_T=-1$.

%
%
%
%
%

According to Ref.~\cite{PhysRevB.83.104426}, physical phenomena in BNFS can be described through symmetric Heisenberg exchange interactions without invoking the DMI.
We thus neglect the DMI and consider symmetric exchange interaction only.
For the structure depicted in Fig.~\ref{fig:crystal}, the intra-layer spin exchange interaction $\mathcal{H}_{\rm intra}^{(l)}$ within the layer $l$, and the inter-layer spin exchange interaction $\mathcal{H}_{\rm inter}^{(l)-(l+1)}$ between the layer $l$ and $l+1$ are as follows:
\begin{eqnarray}
 \mathcal{H}_{\text{intra}}^{ (l)  }
 &=&   \sum_{\alpha \neq \alpha',\beta}
 \biggl[
 J_1  {\bf S}_{l, \alpha } \cdot {\bf S}_{l, \alpha' }
 + J_2
 {\bf S}_{l, \alpha } \cdot {\bf S}_{l, \beta}
 \biggl],  \label{Eq:in-planeH}
 \\
 \mathcal{H}_{\text{inter}}^{ (l)-(l+1) }
 &=&
  \sum_{\alpha, \beta \neq \beta'}
 \biggl[
 J_4 {\bf S}_{l, \alpha } \cdot {\bf S}_{l+1, \alpha }
 + J_5   {\bf S}_{l, \alpha} \cdot {\bf S}_{l+1, \beta }
 \nonumber \\
 && \phantom{asdfasdfasdfasdf}
 + J_3   {\bf S}_{l, \alpha} \cdot {\bf S}_{l+1, \beta' }
 \biggl],
 \label{Eq:inter-planeH}
\end{eqnarray}
where $J_1, J_2, \cdots, J_5$ are exchange parameters [Fig.~\ref{fig:crystal}], and spin operator ${\bf S}_{l, \alpha}$ represents the magnetic moment at $l$, $\alpha$ site.
$l$ represents plane index, and $\alpha$, $\beta$ represent a position vector within the $ab$ plane.
Strictly speaking, $\alpha$ denotes ${\bf R} + {\bf a}_j$, where ${\bf a}_{j}$ is a lattice vector within each Fe$^{3+}$ triangle, ${\bf R}$ is a reference point.
Also $\beta$ denotes ${\bf R} + {\bf R}_k + {\bf a}_j$, where ${\bf R}_k$ ($k=1, 2, 3$) is a inter-triangle position vector.
When an external magnetic field ${\bf B}$ is applied, there appears the Zeeman interaction,
\begin{equation}\label{Eq:Zeeman}
    \mathcal{H}_{\text{z}}^{(l)} = J_0 \sum_{\alpha} {\bf S}_{l,\alpha} \cdot {\bf B}.
\end{equation}
where $J_0 = 2 \mu_B$, and $\mu_B$ is the Bohr magneton.
We assume that ${\bf B}$ is along the $c$ axis, that is, ${\bf B} = B_z \hat{z}$, where $\hat{z}$ denotes the $c$ axis direction [Fig.~\ref{fig:spinconf}].
Then, the total spin Hamiltonian can be obtained by adding up $l$,
\begin{equation}
 \mathcal{H}_{\text{total}}  = \sum_l \left( \mathcal{H}_{\text{intra}}^{ (l) } + \mathcal{H}_{\text{inter}}^{ (l)-(l+1) }   + \mathcal{H}_{\text{Z}}^{ (l) }  \right). \label{eq:totalH}
\end{equation}

To facilitate subsequent analysis, it is convenient to introduce {\it local} coordinate systems whose coordinate axes vary from atomic site to site and are aligned along the local equilibrium spin directions. 
At the site $l$, $\alpha$, the unit vectors for the local coordinate system are
%
%
\begin{eqnarray}
    \hat{x}'_{l, \alpha} &=&
    \hat{x} \cos \theta_{l, \alpha} \cos \phi_{l,  \alpha} + \hat{y}  \cos \theta_{ l,  \alpha} \sin \phi_{ l,  \alpha} + \hat{z}  \sin \theta_{ l,  \alpha},      \nonumber \\
    \hat{y}'_{ l,  \alpha} &=&
      -  \hat{x}  \sin  \phi_{ l,  \alpha} + \hat{y}   \cos \phi_{l,   \alpha},
\label{eq:local-coordinate} \\
    \hat{z}'_{ l,  \alpha} &=&
     - \hat{x} \sin \theta_{ l,  \alpha} \cos \phi_{l, \alpha}   -  \hat{y} \sin \theta_{ l, \alpha} \sin \phi_{ l,  \alpha}  + \hat{z}  \cos  \theta_{ l, \alpha} ,
     \nonumber
\end{eqnarray}
%
%
where $\phi_{l, \alpha}$, $\theta_{l, \alpha}$ are, respectively, azimuthal and polar angles of the equilibrium spin direction at $l$, $\alpha$.
We assume that $\theta_{l, \alpha}$ is independent of $l$ and $\alpha$, that is, $\theta_{l,\alpha}=\theta$.
We also assume that $\phi_{l, {\bf R} + {\bf a}_j }$ follows the helical pattern, that is, 
$\phi_{l, {\bf R} + {\bf a}_j } = \epsilon_H  ( \tau l + \epsilon_T j 2 \pi / 3  )$~\cite{PhysRevLett.101.247201}, where $\tau$ is a helical period along $c$ axis.
Values of $\theta$ and $\tau$ will be determined below by minimizing the equilibrium energy.

Applying the Holstein-Primakoff transformation, the spin operators along local coordinate axes are written as $S_{l, \alpha}^{x'} = S - b^\dag_{l, \alpha} b_{l, \alpha}$, $S_{l, \alpha}^{y'} = \sqrt{S/2} ( b^\dag_{l, \alpha} +  b_{l, \alpha} )$, $S_{l, \alpha}^{z'} = i \sqrt{S/2} (b^\dag_{l, \alpha} - b_{l, \alpha})$, where $b$, $b^\dag$ are bosonic annihilation and creation operators, respectively.
Then, the total Hamiltonian can be expanded in powers of $b$, $b^\dag$.
%
%
\begin{eqnarray}
    \mathcal{H}_{\text{total}}
    &=&
    \mathcal{H}_{\text{total}}^{(0)} + \mathcal{H}_{\text{total}}^{(1)} + \mathcal{H}_{\text{total}}^{(2)}
    \nonumber \\
    &&  \phantom{...}
    + \mathcal{O}( \text{3rd order terms in } b, b^\dag),
\end{eqnarray}
%
%
where $\mathcal{H}_{\rm total}^{(n)}$ denotes $n$-th order terms.
First of all,  $\mathcal{H}_{\rm total}^{(0)}$ reads 
%
\begin{eqnarray}
    && \mathcal{H}_{\text{total}}^{(0)}
     =
    \sum_{l}
    \biggl[
     S^2( J_1 + 2 J_2 ) \biggl( - \frac{1}{2} \cos^2 \theta + \sin^2 \theta \biggl)
     \nonumber \\
     &&
     + \sum_{\nu = 0}^2 S^2 J_{3 + \nu} \biggl(  \cos ( \tau + \epsilon_T \varphi_\nu ) \cos^2 \theta + \sin^2 \theta \biggl)
    \biggl],
\end{eqnarray}
%
where $\varphi_0 = 2 \pi / 3$, $\varphi_1 = 0$, $\varphi_2 = 4 \pi / 3$.
Since $\mathcal{H}_{\rm total}^{(0)}$ amounts to the equilibrium energy, we minimize it with respect to $\theta$ and $\tau$.
From $\partial \mathcal{H}_{\text{total}}^{(0)} / \partial \tau = 0$ and $ \partial \mathcal{H}_{\text{total}}^{(0)} / \partial \theta = 0$, one obtains, 
%
 \begin{eqnarray}
     && \phantom{asdfasdfasdf} 
     \sum_{ \nu = 0}^{2}   J_{3 + \nu} \sin  ( \tau + \epsilon_T \varphi_\nu ) = 0,
    \label{Eq:con-tau}
    \\
     && \sin \theta = -   \frac{ J_0 B_z / S}{3 ( J_1 + 2 J_2) + 2 \sum J_{3 + \nu} \bigl( - \cos ( \tau + \epsilon_T \varphi_\nu) +1 \bigl) }.
    \nonumber \\
    \label{Eq:con-th}
 \end{eqnarray}
%
$\tau$ in Eq.~(\ref{Eq:con-tau}) agrees with the helical period along the $c$ axis reported in Ref.~\cite{PhysRevB.83.104426}. 
On the other hand, Eq.~(\ref{Eq:con-th}) indicates that $\theta=0$ when $B_z=0$ and thus equilibrium spins lie within the $ab$ plane.
When $B_z$ is applied, however, the equilibrium spins deviate from the $ab$ plane [Fig. \ref{fig:spinconf}].
Using the measured parameters in Table II~\cite{PhysRevB.83.104426} that will be used henceforth, we predict $\theta=2.2^\circ$ at $B_z = 6.8 \text{ T}$.
This value is similar to the value reported in Ref.~\cite{JPCM2012Choi}.
Also, the first order term is
\begin{eqnarray}
&& \mathcal{H}^{(1)}_{\text{total}} = 
    i \cos \theta  \sqrt{ \frac{S^3}{2} }  \sum_{l}  \biggl[    \sin \theta   \biggl\{ 3(J_1 + 2 J_2)  
\nonumber \\
&&
    + 2 \sum_{\nu=0}^{2} J_{3+\nu} 
     \bigl( - \cos( \tau + \epsilon_T \varphi_{\nu} ) + 1  \bigl)   \biggl\}  
    +  \frac{ J_0 B_z}{S}   \biggl](b_l^\dagger - b_l).
     \nonumber \\
\end{eqnarray}
This square bracket becomes zero for the value of $\theta$ that minimizes $\mathcal{H}^{(0)}_{\text{total}}$.

The next order term is $\mathcal{H}_{\rm total}^{(2)}$. 
In order to analyze $\mathcal{H}_{\rm total}^{(2)}$, it is convenient to introduce the Fourier transformed bosonic operator $b_{{\bf k}j}$, which is related to $b_{l,{\bf R}+{\bf a}_j}$ as follow
%
\begin{equation}\label{Eq:FTboson}
 b_{l, {\bf R} + {\bf a}_j } = \frac{1}{\sqrt{N}} \sum_{\bf k} \exp \bigl[ i {\bf k} \cdot ( l a_z \hat{z} + {\bf R} ) \bigl] b_{ {\bf k} j},
\end{equation}
%
where $N$ is a number of layers, and $a_z$ is the inter-layer spacing.
In terms of the Fourier transformed bosonic operators, one obtains
%
\begin{equation}
 \mathcal{H}^{(2)}_{\text{total} }
   = \sum_{\bf k} \sum_{j, j'}
   c_{{\bf k}  j}^\dag
   \begin{pmatrix}
    \alpha_{{\bf k}, j j'} & \beta_{{\bf k}, j j'} \\
    \beta_{ {\bf k}, j j'}  & \delta_{{\bf k}, j j'}
   \end{pmatrix}
    c_{{\bf k}  j'}, \label{Eq:6x6H}
\end{equation}
%
where $c_{{\bf k}  j} = \begin{pmatrix} b_{{\bf k}  j} & b_{\bar{ {\bf k}}  j}^\dag \end{pmatrix}^T $ and $\alpha_{{\bf k}, j'j} = \delta_{\bar{ {\bf k}}, j j'}$,
$\beta_{{\bf k}, j' j}  = \beta_{\bar{ {\bf k}}, j j'}$.
Here, $\bar{{\bf k}}$ denotes $-{\bf k}$, and the values of the $\alpha_{{\bf k},j j'}$, $\beta_{{\bf k},j j'}$ are given in Appendix. 
With these values, it is straightforward to verify that Eq.~\eqref{Eq:6x6H} is hermitian.
The equation of motion approach~\cite{doi:10.1063/1.523549,doi:10.1063/1.530338} is a commonly used technique to obtain eigenvalues of the bosonic quadratic Hamiltonian [Eq.~\eqref{Eq:6x6H}]. 
To utilize this approach, we transform Eq.~\eqref{Eq:6x6H} into a standard form of boson quadratic Hamiltonian in Ref.~\cite{doi:10.1063/1.530338} by extending the boson basis $c_{{\bf k} j}$ into $d_{ {\bf k} j} = \begin{pmatrix} b_{ {\bf k} j} & b_{\bar{{\bf k}} j } & b_{ {\bf k} j}^\dag & b_{\bar{{\bf k}} j }^\dag \end{pmatrix}^T$.
Then, Eq.~\eqref{Eq:6x6H} can be rewritten as follows:
%
\begin{equation}
  \mathcal{H}^{(2)}_{\text{total} }
   =
  \frac{1}{2} \sum_{\bf k} \sum_{j, j'} d^\dag_{{\bf k} j}
  \begin{pmatrix}
      A^T_{{\bf k}, j j'} & B_{{\bf k}, j j'} \\
      B_{{\bf k}, j j'}^* & A_{{\bf k}, j j'}
  \end{pmatrix}
  d_{{\bf k} j'}, \label{Eq:12x12H}
\end{equation}
%
which is in the standard form of the boson quadratic Hamiltonian.
Here, $6\times6$ matrices $A_{\bf k}$, $B_{\bf k}$ are
%
\begin{equation}
 A_{{\bf k} } =
 \begin{pmatrix}
  \alpha_{{\bf k} }^T & 0  \\
  0 & \delta_{ {\bf k}  }
 \end{pmatrix},
 \phantom{...}
 B_{{\bf k} } =
 \begin{pmatrix}
  0 & \beta_{{\bf k} }  \\
  \beta_{\bar{{\bf k}} } & 0
 \end{pmatrix},
\end{equation}
%
where $\alpha_{\bf k}$ and $\beta_{\bf k}$ are $3\times 3$ matrices with $\alpha_{{\bf k}jj'}$ and $\beta_{{\bf k}jj'}$ as their matrix elements, respectively.
Then, one obtains the following associated matrix~\cite{doi:10.1063/1.530338} $M_{\bf k}$ 
%
\begin{eqnarray}
 M_{ {\bf k} } &=&
  \begin{pmatrix}
      A_{{\bf k}} & - B_{{\bf k}}^\dag  \\
      B_{{\bf k}} & - A_{{\bf k}}^*
  \end{pmatrix}
  \nonumber \\
  &=&
  \begin{pmatrix}
    \delta_{ \bar{{\bf k}} } & 0 & 0 & - \beta_{\bar{{\bf k}}}^\dag \\
    0 & \delta_{ {\bf k} } & - \beta_{{\bf k}}^\dag & 0 \\
    0 & \beta_{{\bf k}} & - \alpha_{{\bf k}}^\dag & 0 \\
    \beta_{\bar{{\bf k}}} & 0 & 0 & - \alpha_{\bar{{\bf k}}}^\dag
  \end{pmatrix}. \label{Eq:M}
\end{eqnarray}
%
The eigenvalues of this $12\times 12$ matrix $M_{\rm k}$ consists of
$E({\bf k})$, $E(-{\bf k})$, $-E({\bf k})$, and $-E(-{\bf k})$ for three spin wave branches.

\section{\label{sec:result}Result and Discussion}

\begin{figure}[t!]
     \subfigure{\label{fig:spinwave}}
     \subfigure{\label{fig:magnonex}}
     \subfigure{\label{fig:fullband}}
     \subfigure{\label{fig:gapdifference}}
  \includegraphics[width=8.5cm]{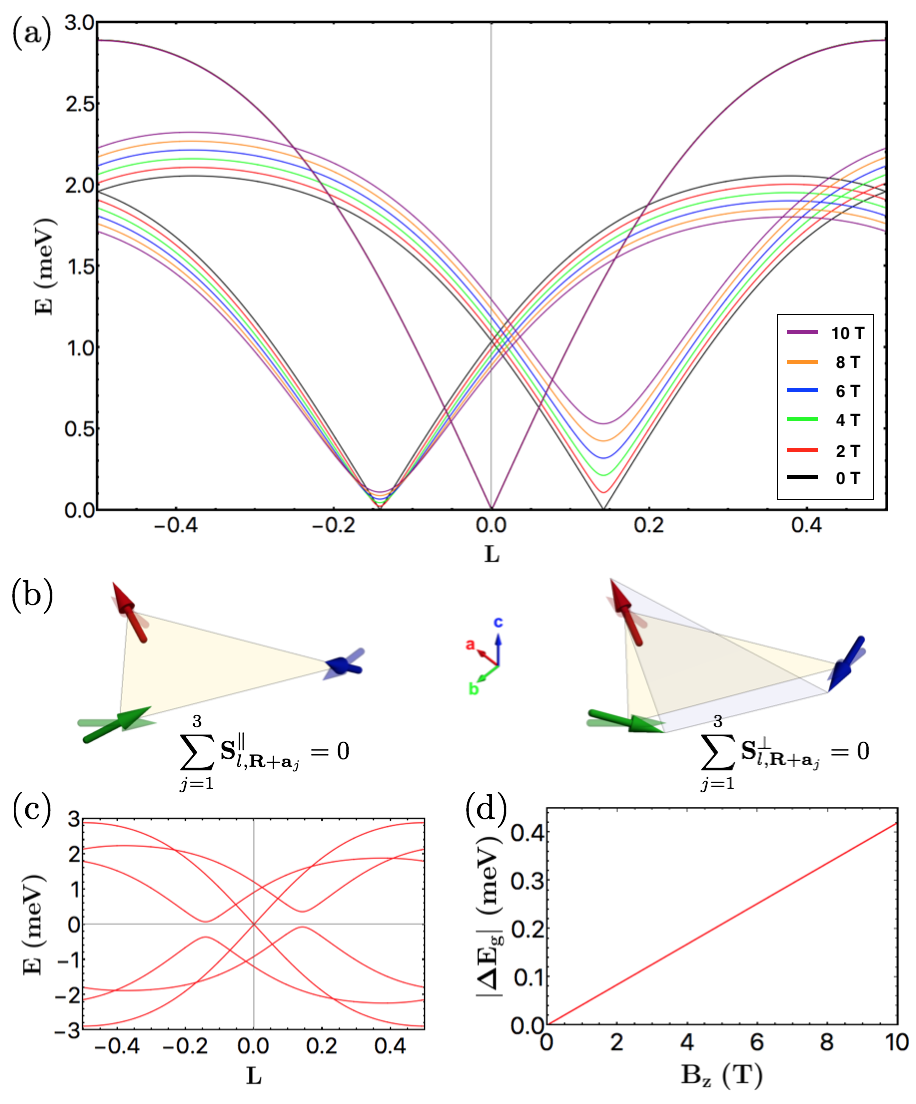}
\caption{
(a) Spin wave dispersion relation with the external magnetic field along the $c$ axis. 
The horizontal axis $\text{L}$ is $\text{L} =  {\bf k} \cdot {\bf a} / 2 \pi$, where ${\bf k}$ is parallel to the $c$ axis and ${\bf a}$ is the lattice vector along the $c$ axis. 
When an external magnetic field ${\bf B}=B_z \hat{\bf z}$ is turned on, the gap asymmetry occurs.
(b) Schematic illustrations of spin wave excitation without ${\bf B}$. For the $c$-mode excitation (left), the summation of the three spins in a Fe$^{+3}$ triangle vanishes within the plane and survives only along the out-of-plane direction.
For the $w_1$- and $w_2$-modes (right), on the other hand, the summation of the three spins in a Fe$^{+3}$ triangle vanishes along the $c$ axis and survives within the plane.
The transparent arrows denote equilibrium spins. 
Here, $\parallel$ ($\perp$) denotes in-plane (out-of-plane) components.
(c) Eigen value spectrum of $M_{{\bf k}}$ for $B_z=6.8$ T.
It shows that the positive and negative energies are origin symmetry.
(d) The energy gap difference between the $w_1$- and $w_2$-modes as a function of $B_z$.
}
\end{figure}

We investigate the spin wave dispersion of BNFS by numerical calculation.
The spin wave dispersion as a function of ${\rm L}=k a_z/2\pi$, and ${\bf B} = B_z \hat{z}$ is shown in Fig.~\ref{fig:spinwave}.
Here we assume that ${\bf k} = (0, 0, k)$.
The different colors indicate different values of $B_z$.
For ${\bf B} = 0 $ (black solid lines), there are three branches of spin wave excitations.
Each of them becomes gapless at $\text{L} = 0$ ($c$-mode), $\text{L} = + \tau / 2 \pi $ ($w_1$-mode), and $\text{L} = - \tau / 2\pi $ ($w_2$-mode), where $\tau/ 2 \pi \simeq 0.14$.
Note that for ${\bf B}={\bf 0}$, $E_{c}({\bf k})=E_{c}(-{\bf k})$ and $E_{w_1}({\bf k}) = E_{w_2}(- {\bf k})$.
Thus the dispersion relations are symmetric. 
As $B_z$ increases from $0$, $E_{c}({\bf k})$ remains essentially unchanged, but $E_{w_1}({\bf k})$ and $E_{w_2}({\bf k})$ are progressively modified.
For both $E_{w_1}({\bf k})$ and $E_{w_2}({\bf k})$, gapless points disappear and are replaced by quadratic dispersions.
Note that the resulting energy gap is significantly bigger for the $w_1$-mode than for the $w_2$-mode.
Thus the relation $E_{w_1}({\bf k}) = E_{w_2}(- {\bf k})$ becomes broken and the dispersions for the $w_1$- and $w_2$-modes become asymmetric, acquiring the non-reciprocity. 
In addition, we remark that the sign of the non-reciprocity [Fig.~\ref{fig:spinwave}] can be reversed when the sign of $B_z$ is reversed. 
The sign of the non-reciprocity can be also reversed when the sign of the magnetic chirality $\epsilon_T \epsilon_H$ is reversed, although the magnetic chirality reversal is difficult to realize in experiments because this reversal requires energy costs.

To understand this result, it is useful to consider the nature of spin wave ``vibrations''.
Figure~\ref{fig:magnonex} shows schematically the spin vibration patterns for the $c$-mode excitation (left), and the $w_{1/2}$-mode excitation within a Fe$^{+3}$ triangle.
In the $c$-mode, all spins vibrate without altering their net in-plane component, hence $\sum^3_{j=1} \delta {\bf S}^{\parallel}_{l, {\bf R} + {\bf a}_j} = 0$ within the triangle. 
Here, $\parallel$ denotes in-plane components.
For this mode, the system has the rotation symmetry around the $c$ axis regardless of whether $B_z$ is applied. Thus $E_c({\bf k}={\bf 0})$ for arbitrary $B_z$, since this particular mode amounts to the Goldstone mode for the symmetry. 
In case of $w_{1/2}$-modes, on the other hand, the spins vibrate without alternating their net $c$ component, hence $\sum^3_{j=1} \delta {\bf S}^{\perp}_{l, {\bf R} + {\bf a}_j} = 0$ within the triangle. 
Here, $\perp$ denotes out-of-plane components.

The blue plane in Fig.~\ref{fig:magnonex}, which is defined by connecting the end points of the vibrating spins, shows the out-of-plane vibration clearly.
In the $w_1$- and $w_2$-modes, the normal vector to the blue plane precess around the $c$ axis anticlockwise and clockwise, respectively.
If ${\bf B} = 0$, the anticlockwise and clockwise precessions share the same vibration frequencies, resulting in $E_{w_1}({\bf k})=E_{w_2}(-{\bf k})$ [black solid line in Fig.~\ref{fig:spinwave}].
For $B_z \neq {\bf 0}$, on the other hand, the field itself tends to induce the precession of the normal vector in one particular direction, thus introducing the different between the anticlockwise and clockwise precession frequencies.
This explains the non-reciprocity, $E_{w_1}({\bf k})\ne E_{w_2}(-{\bf k})$ in the $w_1$- and $w_2$-modes.
Figure~\ref{fig:gapdifference} shows that the difference $E_{w_1}( {\bf k}=     \tau  /a_z \hat{\bf z} ) - E_{w_2}(- {\bf k}=  -  \tau  /a_z \hat{\bf z} )$ between the energy gaps of the $w_1$- and $w_2$-modes increases with increasing $B_z$.
For $B_z = 6.8 \text{ T}$, the energy gaps for the $w_1$- and $w_2$ modes are $ 0.36 \text{ meV}$, and $0.07 \text{ meV}$, respectively.
Then, one obtains the gap size difference of $0.29 \text{ meV}$.
%

In some respect, our result is similar to Ref.~\cite{PhysRevLett.119.047201} that reports the non-reciprocal spin waves in a chiral antiferromagnet $\alpha$-Cu$_2$V$_2$O$_7$ with ${\bf B}$.
However, there are distinctions between BNFS and $\alpha$-Cu$_2$V$_2$O$_7$ systems.
In BNFS, the chiral antiferromagnetic order arises from competition between symmetric exchange interactions whereas in $\alpha$-Cu$_2$V$_2$O$_7$, it arises from the DMI.
Another important difference is the spin wave propagation direction.
In BNFS, spin waves propagating parallel to the external magnetic field are non-reciprocal where in $\alpha$-Cu$_2$V$_2$O$_7$, spin waves propagating {\it perpendicular} to the external magnetic field are non-reciprocal.

Let us investigate the structure of Eq.~\eqref{Eq:M} more closely to better understand reciprocal (non-reciprocal) spin wave without (with) $B_z$.
First of all, we remark that eigenvalues of $M_{\bf k}$ are real even though $M_{\bf k}$ is not hermitian. Given this information, we can understand the reciprocity (non-reciprocity) as follows.
The characteristic equation $\det \bigl[ M_{ {\bf k}  } - x \textbf{I}  \bigl] = 0$ is rewritten as
\begin{equation}
 \begin{vmatrix}
     \delta_{\bar{{\bf k}}}  - x  \textbf{I}  & - \beta_{\bar{{\bf k}}}^\dag \\
     \beta_{\bar{{\bf k}}}    & - \alpha_{\bar{{\bf k}}}^\dag - x  \textbf{I} 
 \end{vmatrix} 
 \times 
 \begin{vmatrix}
    \delta_{{\bf k}}  - x  \textbf{I}  & - \beta_{{\bf k}}^\dag \\
     \beta_{{\bf k}}    & - \alpha_{{\bf k}}^\dag - x  \textbf{I}
 \end{vmatrix} = 0, 
 \label{Eq:EigM}
\end{equation}
where $\textbf{I}$ is $3\times3$ identity matrix, $x$ is a real eigenvalue of $M_{\bf k}$, and the first (second) determinant in the left hand side is for $- {\bf k}$ ($+ {\bf k}$).
When $B_z=0$, matrix elements for ${\bf k}$ and $\bar{\bf k}$ are related (see Appendix) as follows:
\begin{equation}\label{eq:reciM}
 \begin{pmatrix}
      \delta_{ {\bf k} } & - \beta_{ {\bf k} }^\dag \\
      \beta_{{\bf k} } & - \alpha_{ {\bf k} }^\dag
 \end{pmatrix}^*
 =
 \begin{pmatrix}
      \delta_{ \bar{{\bf k}} } & - \beta_{ \bar{{\bf k}} }^\dag \\
      \beta_{ \bar{{\bf k}} } & - \alpha_{ \bar{{\bf k}} }^\dag
 \end{pmatrix}.
\end{equation}
Then comparing the first and second determinants, and recalling that $x$ is real, one finds that the first (for $-{\bf k}$) and the second (for ${\bf k}$) produce the same eigenvalues. 
Thus the spin wave spectrum is reciprocal. 
When $B_z\neq 0$, on the other hand, the relation in Eq.~\eqref{eq:reciM} breaks down since 
$\alpha^*_{{\bf k}} \neq \alpha_{\bar{{\bf k}}}$, $\delta^*_{{\bf k}} \neq \delta_{\bar{{\bf k}}}$ while $\beta^*_{{\bf k}} = \beta_{\bar{{\bf k}}}$.
The explicit expression for $\alpha_{{\bf k}}$ ($= \delta^{T}_{\bar{{\bf k}}}$) is given in Eq.~\eqref{Eq:alpha}, and the last term of this expression breaks the relation.
Therefore, the low energy excitation spin wave spectrum may become non-reciprocal, $E(+{\bf k}) \neq E(-{\bf k})$.

%
%
%
\begin{figure}[t!] 
     \subfigure{\label{fig:4a}}
     \subfigure{\label{fig:4b}}
     \subfigure{\label{fig:4c}}
     \subfigure{\label{fig:4d}}
  \includegraphics[width=8.5cm]{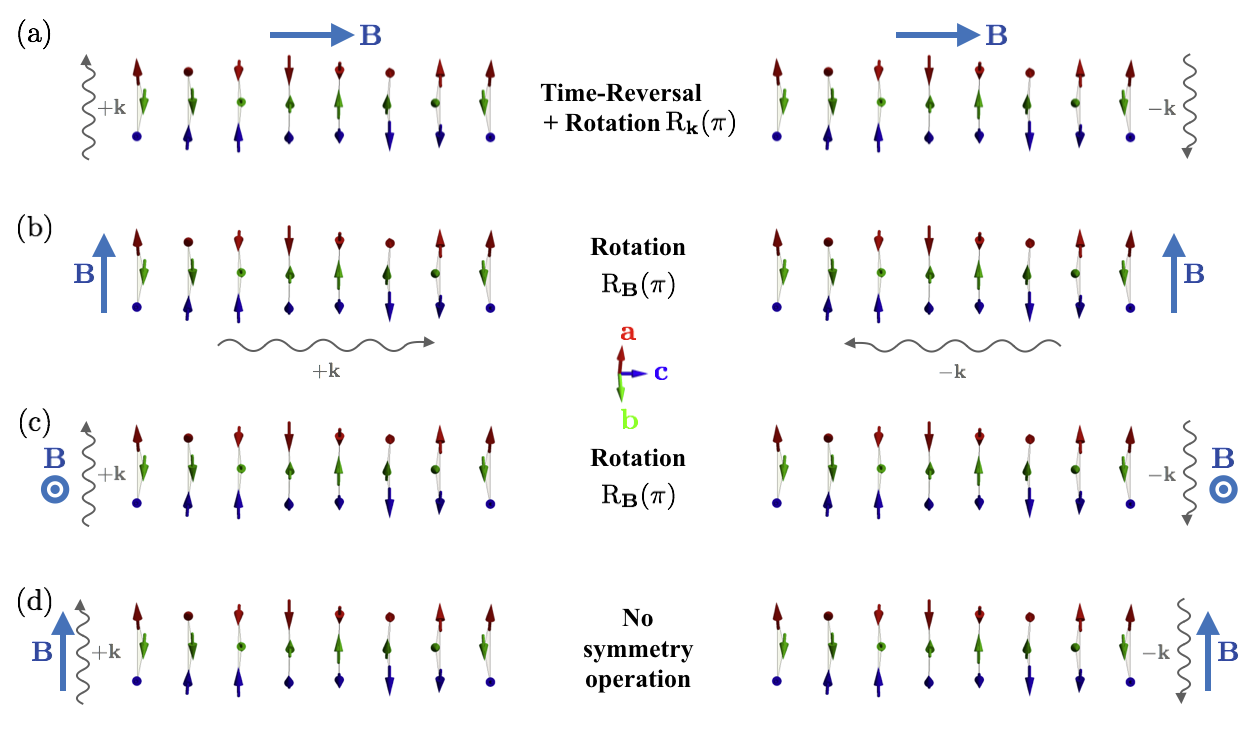}
\caption
{Connection between spin waves with $+{\bf k}$ and $-{\bf k}$ through symmetry operations.
(a) When ${\bf B}$ (or a magnetization ${\bf M}$) is applied along the $c$-direction, a spin wave propagating along the $a$-axis (or the $b$-axis) becomes reciprocal, because the combined symmetry operation of the time-reversal and rotation around the ${\bf k}$ direction links the two spin wave cases.
(b), (c) Because spin waves with $\pm {\bf k}$, which are perpendicular to the ${\bf B}$ direction, can be connected by a rotation around ${\bf B}$ (or ${\bf M}$) direction, we obtain reciprocal spin waves.
(d) When ${\bf k}$ and ${\bf B}$ (or ${\bf M}$), which are in the $ab$ plane, are parallel to each other, there does not exist any symmetry operation that links the two spin waves.
Hence the spin wave dispersion in this case can becomes non-reciprocal.
The rotations $\text{R}_{\bf k}(\pi)$, and $\text{R}_{{\bf B}}(\pi)$ rotate around ${\bf k}$, and ${\bf B}$ directions, respectively. 
Note that ${\bf k}$ in (a), and ${\bf B}$ (or ${\bf M}$) in (b), (c), and (d) are parallel to the $a$-axis (or the $b$-axis).
}
\label{fig:4}
\end{figure}
%
%

So far our analysis of BNFS has focused on the non-reciprocity from symmetric exchange interactions [Eq.~\eqref{Eq:in-planeH} and \eqref{Eq:inter-planeH}] and neglected the DMI.
To be strict, the DMI may also exist in BNFS since it is noncentrosymmetric.
According to \cite{PhysRevB.84.104405}, the energy scale of the DMI is three orders of magnitude smaller than $J_1$, and about two orders of magnitude smaller than $J_2$, $J_3$, $J_4$, $J_5$. 
Although such small DMI can generate observable effects such as energy gap opening~\cite{PhysRevB.84.104405} to the $w_1$- and $w_2$-modes, it can not significantly affect the degree of the non-reciprocity (energy gap difference between the $w_1$- and $w_2$-modes) simply because the DMI energy scale is much smaller than those of $J$'s.
%

Our examination of the non-reciprocal spin waves in BNFS focused on the case when ${\bf k}$ and ${\bf B}$ are parallel to the $c$-axis [Fig.~\ref{fig:spinconf}].
Figure~\ref{fig:4} shows possible other configurations of ${\bf k}$ and ${\bf B}$, which are not examined in this paper.
For the three cases with ${\bf k}$ and ${\bf B}$ perpendicular to each other [depicted in Figs.~\ref{fig:4a}, \ref{fig:4b}, and \ref{fig:4c}], one can show by using simple symmetry argument~\cite{Cheong:2018kg} that the non-reciprocity is not possible.
First, for the case in Fig.~\ref{fig:4a}, the spin waves propagating along the $(+a)$-axis and $(-a)$-axis becomes reciprocal, because a combination of the time-reversal and rotation operations can connect the two spin-wave propagating directions.
The same symmetry argument applies to the spin waves propagating along $(+b)$-axis and $(-b)$-axis.
However this argument does not apply to the spin waves propagating along the direction which deviates from the $\pm {\bf a}$ or $\pm {\bf b}$ directions, even though the propagation direction lies within the $ab$ plane.
%
%
\begin{figure}[t!]
     \subfigure{\label{fig:5a}}
     \subfigure{\label{fig:5b}} 
  \includegraphics[width=8.5cm]{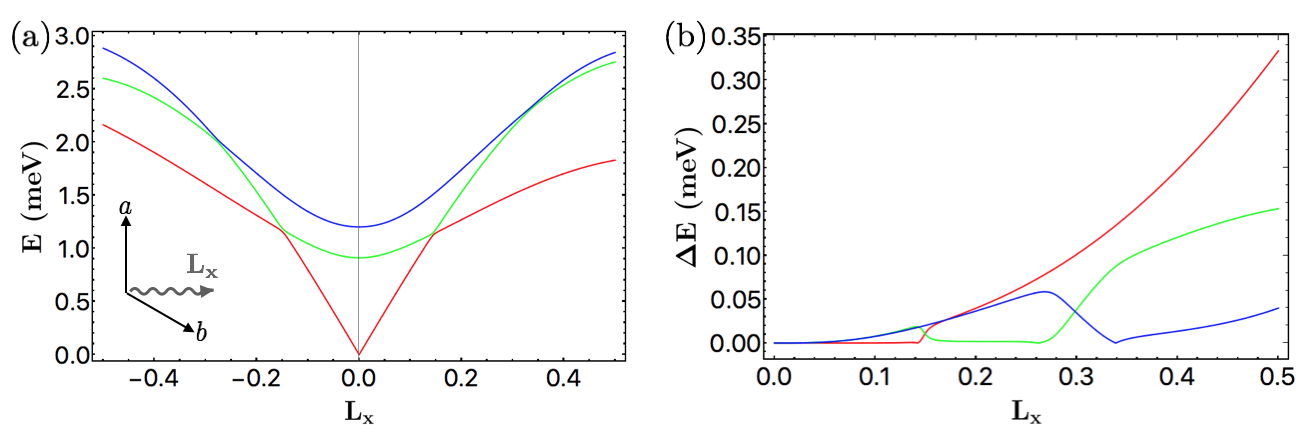}
\caption
{(a) Asymmetric spin wave dispersion relation for $B_z = 6.8 \text{ T}$.
Inset indicates the direction of $\text{L}_x$, which lies within the $ab$ plane.
(b) Energy difference $\Delta E(\text{L}_x) = E(+\text{L}_x) - E(-\text{L}_x)$.
Colors are introduced to identify the corresponding dispersions in (a). 
}
\label{fig:5}
\end{figure}
%
%
For example, when ${\bf k}$ is parallel to $({\bf a} + 2 {\bf b})/\sqrt{3}$, the spin wave dispersion [Fig.~\ref{fig:5a}] shows much weaker non-reciprocity than Fig.~\ref{fig:spinwave}.
Here an external magnetic field is along the $c$ axis.
In this case, each spin wave branch is asymmetric.
In order to show asymmetric spin wave clearly, we provide the energy difference between $+\text{L}_x$ and $- \text{L}_x$ as function of $\text{L}_x$ [Fig.~\ref{fig:5b}].
Here the non-reciprocal effect increases with increasing spin wave propagation vector.
However, this effect is strongly suppressed in the vicinity of ${\text L}_x = 0$.
Thus, we expect it to be weak at least for the long wave length spin wave.
Second, when ${\bf B}$ is applied in the $ab$ plane [Figs.~\ref{fig:4b} and \ref{fig:4c}], the actual calculation of spin wave excitations become complicated since the ground state spin configuration is not known for this case.
But the symmetry analysis may be still possible if a net magnetization in BNFS is parallel to ${\bf B}$.
Then, spin waves in Figs.~\ref{fig:4b} and \ref{fig:4c} are reciprocal, since spin waves with $\pm {\bf k}$ can be connected by a rotation around the ${\bf B}$ direction.
We remark that this symmetry argument becomes exact when ${\bf B}$ is parallel to $\pm a$-axis or $\pm b$-axis.
Finally, we consider the case in Fig.~\ref{fig:4d}, where ${\bf k}$, ${\bf B}$ are parallel to each other, and these are within the $ab$ plane.
In this case, two spin waves with ${\bf k} \parallel {\bf B}$ and $-{\bf k} \parallel {\bf B}$ cannot be connected by any operations, that is, this spin wave can become non-reciprocal.
Therefore, the spin wave in BNFS may be non-reciprocal (reciprocal) when ${\bf k}$ and ${\bf B}$ are parallel (perpendicular) to each other.

\section{\label{sec:summary}Summary}
In summary, we have shown theoretically that the spin wave in a chiral antiferromagnet BNFS becomes non-reciprocal when an external magnetic field is applied along the $c$-axis.
Unlike other chiral ferromagnets or chiral antiferromagnets, where DMI is crucial for the non-reciprocity, the non-reciprocity in BNFS, which has very small DMI, can arise purely from the competition between symmetric exchange interactions and an external magnetic field.
Thus our work demonstrates that the DMI is not crucial for non-reciprocal spin waves.
Our work also widens material choice for non-reciprocal spin waves.

\begin{acknowledgments}
HWL and SC were  supported by the National Research Foundation of Korea (NRF)
grant funded by the Korea government (MSIT) (No.2018R1A5A6075964).
SWC was supported by the DOE under Grant No. DOE: DE-FG02-07ER46382.
\end{acknowledgments}

\revappendix*
\section{Components of Eq. \eqref{Eq:6x6H}}
$\alpha_{{\bf k}}$, $\beta_{{\bf k}}$ are presented as below:
\begin{eqnarray}
 \alpha_{{\bf k}}
 &=&
  \biggl[
   - \frac{1}{2} J_0 B_z \sin \theta + \frac{1}{2} ( J_1 + 2 J_2 )S ( 1 - 3 \sin^3 \theta)
  \nonumber \\
  &&
   - \sum_{\nu = 0}^{2} J_{4 + \nu} S
   \biggl\{  \cos( \tau + \epsilon_T \varphi_\nu) \cos^2 \theta + \sin^2 \theta \biggl\}
  \biggl] {\bf I}
  \nonumber \\
  &&
  + J_1 S
  \biggl[
     \frac{1}{8} ( 1 - 3 \sin^2 \theta)  \phantom{.} \Sigma_{3 x}
     -  \frac{ \sqrt{3}}{4} \epsilon_H \epsilon_T \phantom{.} \sin \theta  \phantom{.} \Sigma_{3y}
  \biggl]
  \nonumber \\
  &&
  +J_2 S
  \biggl[
  \frac{1}{8} ( 1 - 3 \sin^2 \theta)  \phantom{.} \mathcal{A}^{+}_{3 {\bf k}}
  - \frac{ \sqrt{3}  }{4} \epsilon_H \epsilon_T \phantom{.} \sin \theta  \phantom{.} \mathcal{A}^{-}_{3 {\bf k}}
  \biggl]
  \nonumber \\
  &&
  + \frac{S}{4} ( 1 + \sin^2 \theta) \phantom{.} \mathcal{B}^{+}_{3 {\bf k}}
  + \frac{S}{4} \cos^2 \theta \phantom{.}  \mathcal{C}_{3 {\bf k}}
  - \frac{S}{2} \epsilon_H \phantom{.} \sin \theta \phantom{.}  \mathcal{B}^{-}_{3 {\bf k}}, \label{Eq:alpha}
  \nonumber \\
\end{eqnarray}
\begin{eqnarray}
  \beta_{{\bf k}}
  &=&
  - \frac{3}{8} S ( 1 - \sin^2 \theta) ( J_1  \Sigma_{3x} + J_2 \mathcal{A}_{3 {\bf k} }^{+} )
  \nonumber \\
  && \phantom{asdfasdf}
  + \frac{1}{4} S ( 1 - \sin^2 \theta) ( \mathcal{B}_{3 {\bf k}}^{+} - \mathcal{C}_{ 3 {\bf k} } ) S ,
\end{eqnarray}
where
\begin{equation}
 \Sigma_{3x} =
 \begin{pmatrix}
     0 & 1 & 1 \\
     1 & 0 & 1 \\
     1 & 1 & 0
 \end{pmatrix}
 , \phantom{.}
  \Sigma_{3y} =
 \begin{pmatrix}
     0 & -i & i \\
     i & 0 & -i \\
     -i & i & 0
 \end{pmatrix},
\end{equation}

\begin{eqnarray}
 ( \mathcal{A}^+_{3 {\bf k}} )_{lm}
 &=&
 \bigl| \hat{e}_l \times \hat{e}_m \bigl| \phantom{.} ( e^{ i {\bf k} \cdot {\bf R}_l} + e^{ - i {\bf k} \cdot {\bf R}_m }),
 \label{eq:A+}
\end{eqnarray}
\begin{eqnarray}
  ( \mathcal{A}^-_{3 {\bf k}} )_{lm}
 &=&
 i ( \hat{e}_l \times \hat{e}_m ) \cdot \hat{e}_n \phantom{.} ( e^{ i {\bf k} \cdot {\bf R}_l} + e^{ - i {\bf k} \cdot {\bf R}_m }),
 \label{eq:A-}
\end{eqnarray}

\begin{eqnarray}
 ( \mathcal{B}^+_{3 {\bf k}} )_{lm}
 &=&
 2 \delta_{lm}  J_4' \cos k_z a_z + \bigl| \hat{e}_l \times \hat{e}_m \bigl| \phantom{.} J_+' \cos k_z a_z
 \nonumber \\
 && \phantom{as}
 + i ( \hat{e}_l \times \hat{e}_m ) \cdot \hat{e}_n \phantom{.} J_-' \sin k_z a_z,
 \end{eqnarray}
 \begin{eqnarray}
 ( \mathcal{B}^-_{3 {\bf k}} )_{lm}
 &=&
 2 \delta_{lm}  J_4'' \sin k_z a_z + \bigl| \hat{e}_l \times \hat{e}_m \bigl| \phantom{.} J_+'' \sin k_z a_z
 \nonumber \\
 && \phantom{as}
 - i ( \hat{e}_l \times \hat{e}_m ) \cdot \hat{e}_n \phantom{.} J_-'' \cos k_z a_z,
 \end{eqnarray}
and
 \begin{eqnarray}
 ( \mathcal{C}_{3 {\bf k}} )_{lm}
 &=&
 2 \delta_{lm}  J_4 \cos k_z a_z + \bigl| \hat{e}_l \times \hat{e}_m \bigl| \phantom{.} J_+ \cos k_z a_z
 \nonumber \\
 && \phantom{as}
 + i ( \hat{e}_l \times \hat{e}_m ) \cdot \hat{e}_n \phantom{.} J_- \sin k_z a_z.
\end{eqnarray}
Here, $n \neq l, m$, and $l,m = 1, 2, 3$.
$\hat{e}_{l}$ is a unit vector.
$J_\pm  = J_5 \pm i  J_6 $, $J_{4 + \nu}' = J_{4 + \nu} \cos ( \tau + \epsilon_T \varphi_\nu )$, and $J_{4 + \nu}'' = J_{4 + \nu} \sin ( \tau + \epsilon_T \varphi_\nu )$. 
Since we assume that ${\bf k} = (0, 0, k)$, Eqs. \eqref{eq:A+}, \eqref{eq:A-} can be written as $\mathcal{A}_{3 {\bf k} }^{+} = 2 \Sigma_{3x}$, $\mathcal{A}_{3 {\bf k} }^{-} = 2 \Sigma_{3y}$.

\nocite{*}
\bibliography{ref}

\end{document}